\documentclass[conference]{IEEEtran}
\usepackage[noadjust]{cite}

\ifCLASSINFOpdf
   \usepackage[pdftex]{graphicx}
\else
   \usepackage[dvips]{graphicx}
\fi
\usepackage{amsfonts}
\usepackage{bm}
\usepackage{supertabular}
\usepackage{longtable}
\usepackage[cmex10]{amsmath}
\usepackage{fixltx2e}
\usepackage{url}

\begin{document}
\title{Constructions of Pure Asymmetric Quantum Alternant Codes Based on Subclasses of Alternant Codes}
%
%
%
\author{\IEEEauthorblockN{Jihao Fan}
\IEEEauthorblockA{Department
of Computer Science and  Engineering\\
Southeast University\\
Nanjing, Jiangsu 211189, China\\
Email: fanjh12@seu.edu.cn}
\and
\IEEEauthorblockN{Hanwu~Chen}
\IEEEauthorblockA{Department
of Computer Science and   Engineering\\
Southeast University\\
Nanjing, Jiangsu 211189, China\\
Email: hw\_chen@seu.edu.cn}
}

%
%

\markboth{DECEMBER ~2013}%
{Shell \MakeLowercase{\textit{et al.}}: Bare Demo of IEEEtran.cls for Journals}
%



\maketitle

\begin{abstract}
In this paper, we construct
asymmetric quantum error-correcting codes(AQCs) based on subclasses of Alternant codes. Firstly, We propose a new subclass of Alternant codes which can attain the classical Gilbert-Varshamov bound to construct AQCs. It is shown that when $d_x=2$, $Z$-parts of the AQCs can attain the classical Gilbert-Varshamov bound. Then we construct AQCs based on a famous subclass of Alternant codes called Goppa codes. As an illustrative example, we get three $[[55,6,19/4]],[[55,10,19/3]],[[55,15,19/2]]$ AQCs from the well known $[55,16,19]$ binary Goppa code. At last, we get asymptotically good binary expansions of asymmetric quantum GRS codes, which are quantum generalizations of Retter's classical results. All the AQCs constructed in this paper are pure.
\end{abstract}


%
\IEEEpeerreviewmaketitle

\section{Introduction}

\newtheorem{definition}{\bfseries\upshape Definition}[section]
\newtheorem{theorem}[definition]{\bfseries\upshape Theorem}
\newtheorem{lemma}[definition]{\bfseries\upshape Lemma}
\newtheorem{corollary}[definition]{\bfseries\upshape Corollary}
\newtheorem{example}[definition]{\bfseries\upshape Example}
\newtheorem{proposition}[definition]{\bfseries\upshape Proposition}
%
%
%
%
In many quantum mechanical systems the mechanisms for
the occurrence of bit flip and phase flip errors are quite
different.
Recently, several papers argue that in most of
the known quantum computing models, the
phase-flip errors ($Z$-type errors) happen more frequently than the
bit-flip errors ($X$-type errors) and other types of errors. And the asymmetry is large in
general \cite{evans2007error}.
Motivated by this phenomena,
asymmetric quantum error-correcting codes (AQCs) are
designed to adjust this asymmetry, which may have more flexbility than
general quantum error-correcting codes (QECs).

Steane first stated the importance of AQCs in \cite{steane1996multiple}.
Some recent progress is given in
\cite{aliferis2008fault,evans2007error,brooks2013fault}.
Sarvepalli $et\ al.$ constructed AQCs using a combination of
BCH and finite geometry
LDPC codes in \cite{sarvepalli2009asymmetric}.
A more comprehensive
characterization of AQCs was given by Wang $et\ al.$ which unified
the nonadditive AQCs as well \cite{wang2010asymmetric}.
Ezerman $et\ al.$ \cite{frederic2013csslike} proposed so-called CSS-like
constructions based on pairs of nested subfield linear codes. They also
used  nested codes (such as BCH codes, circulant codes, etc.)
over $\mathbb{F}_4$ to construct AQCs in their earlier
work \cite{ezerman2011additive}. The asymmetry was introduced into topological quantum codes in \cite{fowler2013analytic}.

Alternant codes are a very large family of linear
error-correcting codes. Many
interesting and famous subclasses of Alternant codes have been obtained, for instance,
BCH codes, Goppa codes, etc.
There exist long Alternant codes meeting the Gilbert-Varshamov bound.
BCH codes and GRS codes have been widely used to construct QECs \cite{ketkar2006nonbinary} and AQCs \cite{ezerman2013pure,sarvepalli2009asymmetric}. However,
other subclasses of Alternant codes have received less attention. And there is
an important problem that whether existing asymptotically good quantum Alternant
codes could attain the quantum Gilbert-Varshamov bound. Inspired by these,
we carry out the construction of asymmetric quantum Alternant codes.

\section{Preliminaries}
Let $p$ be a prime number and $q$ a power of $p$, i.e., $q=p^r$ for some $r>0$.
Let $\mathbb{F}_q$ denote the finite field with $q$ elements. The finite field $\mathbb{F}_{q^m}$ is a field extension of degree $m$ of the
field $\mathbb{F}_q$. The trace mapping
${\rm Tr}:\mathbb{F}_{q^m}\rightarrow \mathbb{F}_q$ is given by
${\rm Tr}(a)=a+a^q+\ldots+a^{q^{m-1}}$, for $a\in\mathbb{F}_{q^m}$.
\subsection{Classical Codes}
We review some basic results of GRS codes and Alternant codes firstly.

The Reed-Solomon code  of length $n=q^m-1$(denoted by $\mathcal{RS}(n,\delta)$) is a cyclic code over $\mathbb{F}_{q^m}$ with roots $1$, $\alpha,\ldots,\alpha^{\delta-2}$,
where $\delta$ is an integer, $2\leq\delta\leq n-1$, $\alpha$ is a
primitive element of $\mathbb{F}_{q^m}$.  The parameters
of $\mathcal{RS}(n,\delta)$ are $[n,k,d]_{q^m}$, where $k=n-\delta+1$, $d=\delta$. The parity check matrix of $\mathcal{RS}(n,\delta)$ is given by
\begin{eqnarray}
\label{parity check matrix of general RS codes}
H_{\mathcal{RS}(n,\delta)}=
\left(
\begin{array}{cccc}
1&1&\cdots&1\\
1&\alpha&\cdots&\alpha^{n-1}\\
\vdots&\vdots&\vdots&\vdots\\
1&\alpha^{\delta-2}&\cdots&\alpha^{(n-1)(\delta-2)}
\end{array}
\right).
\end{eqnarray}

GRS codes are  obtained by a further generalization of  RS codes. Let
$\mathbf{a}=(\alpha_1, \alpha_2, \ldots, \alpha_n)$ where the $\alpha_i$
are distinct elements of $\mathbb{F}_{q^m}$, and let
$\mathbf{v}=(v_1, v_2, \ldots, v_n)$ where the $v_i$ are
nonzero elements of $\mathbb{F}_{q^m}$. For any $1\leq k\leq n-1$,
the GRS code $\mathcal{GRS}_k(\mathbf{a},\mathbf{v})$ is defined by
\begin{eqnarray}
\label{definition of GRS codes}
\nonumber
\mathcal{GRS}_k(\mathbf{a},\mathbf{v})=\big\{(v_1F(\alpha_1), v_2F(\alpha_2), \ldots, v_nF(\alpha_n))\ |\\\
 F(x)\in \mathbb{F}_{q^m}[x],\ \deg F(x)<k \big\}.
\end{eqnarray}
The parameters of $\mathcal{GRS}_k(\mathbf{a},\mathbf{v})$ are
$[n,k,n-k+1]_{q^m}$. 
The dual of a GRS code is also a
GRS code, i.e., $\mathcal{GRS}_k(\mathbf{a},\mathbf{v})^\bot=\mathcal{GRS}_{n-k}(\mathbf{a},\mathbf{y})$,
where $\mathbf{y}=(y_1, y_2, \ldots, y_n)$ and $y_i\cdot v_i=1/\prod_{j\neq i}(\alpha_i-\alpha_j)$, for $1\leq i\leq n$. The parity check
matrix of $\mathcal{GRS}_k(\mathbf{a}, \mathbf{v})$ is given by
\begin{equation}
\label{parity check matrix of alternant codes}
H_{\mathcal{GRS}_k(\mathbf{a},\mathbf{v})} =
\left(
\begin{array}{cccc}
y_1&y_2&\cdots&y_n\\
\alpha_1y_1&\alpha_2y_2&\cdots&\alpha_ny_n\\
\vdots&\vdots&\vdots&\vdots\\
\alpha_1^{r-1}y_1&\alpha_2^{r-1}y_2&\cdots&\alpha_n^{r-1}y_n
\end{array}
\right)
\end{equation}
where $r=n-k$.

Both RS codes and GRS codes are MDS codes. The Hamming weight enumerator of any MDS code $[n, k, d]_q$ where $d=n-k+1$  is
completely determined by
\begin{equation}
\label{Weight Distribution of MDS Codes}
A_w = \binom{n}{w}(q-1)\sum_{j=0}^{w-d}(-1)^j \binom{w-1}{j}q^{w-d-j}
\end{equation}
from \cite{macwilliams1977theory}.

Alternant codes are obtained as subfield subcodes of GRS codes. For the notation given above, Alternant code
$\mathcal{A}_r(\mathbf{a}, \mathbf{y})$ is defined as
$\mathcal{A}_r(\mathbf{a}, \mathbf{y})=\mathcal{GRS}_k(\mathbf{a}, \mathbf{v})\ |\ \mathbb{F}_q$. Therefore $\mathcal{A}_r(\mathbf{a}, \mathbf{y})$ has the same parity check matrix as $\mathcal{GRS}_k(\mathbf{a}, \mathbf{v})$.
\subsection{Quantum Error-Correcting codes }
Let $\mathbb{C}$ be the complex number field. For a positive integer $n$,
let $V_n=(\mathbb{C}^q)^{\otimes n}=\mathbb{C}^{q^n}$ be the $n$th tensor
product of $\mathbb{C}^q$.
\begin{definition}
\label{definition of QEC and AQC}
A $q$-ary asymmetric quantum code of length $n$, denoted by $[[n,k,d_z/d_x]]_q$ is a subspace $Q$ of $V_n$ over finite field $F_q$ with dimension $q^k$, which can detect $d_x-1$ qubits of $X$-errors and, at the same time, $d_z-1$ qubits
of $Z$-errors.
\end{definition}

\begin{lemma}[AQCs Constructions\cite{sarvepalli2009asymmetric,wang2010asymmetric}]
\label{AQC Constructions}
Let $C_1$ and $C_2$ denote two classical
linear codes with parameters
$[n,k_1,d_1]_q$ and $[n,k_2,d_2]_q$ such that $C_2^{\bot}\subseteq C_1$. Then there exists an $[[n,k_1+k_2-n,d_z/ d_x]]_q$ AQC, where
$d_z=\text {wt}(C_1\backslash C_2^{\bot})$ and $d_x=\text {wt}(C_2\backslash C_1^{\bot})$.
If $d_z=d_1$ and $d_x=d_2$, then the code is pure.
\end{lemma}

For a given pair $(\delta_x,\delta_z)$ of real numbers and a family $\mathcal{Q}=\big\{[[n^{(i)},k^{(i)},d_z^{(i)}/d_x^{(i)}]]\big\}_{i=1}^{\infty}$ of
asymptotic quantum codes with
\[
\mathop{\lim\inf}_{i\rightarrow\infty}\frac{d_x^{(i)}}{n^{(i)}}\geq \delta_x,\hspace{5mm}
\mathop{\lim\inf}_{i\rightarrow\infty}\frac{d_z^{(i)}}{n^{(i)}}\geq \delta_z
\]
denote the asymptotic quantity as
\[
R_\mathcal{Q}(\delta_x,\delta_z)=\mathop{\lim\sup}_{i\rightarrow\infty}\frac{k^{(i)}}{n^{(i)}}
\]

One of the central
asymptotic problems for quantum codes is to find families $\mathcal{Q}$
of asymptotic quantum codes such that for a fixed pair $(\delta_x,\delta_z)$,
the value $R_\mathcal{Q}(\delta_x,\delta_z)$ is as large as possible. The best known nonconstructive lower bound on $R_\mathcal{Q}(\delta_x,\delta_z)$
can be obtained from \cite{calderbank1998quantum}:
\begin{equation} R_\mathcal{Q}(\delta_x,\delta_z)\geq 1-H(\delta_x)-H(\delta_z)
\end{equation}
where $H(x)=-x\log_2x-(1-x)\log_2(1-x)$ is the binary entropy
function. It is the quantum Gilbert-Varshamov bound for AQCs.

\section{Asymptotically $Z$-parts Good Asymmetric Quantum Alternant Codes}
We take $\mathbf{y}=(y_1, y_2, \ldots, y_n)$ as the encoded codeword of the RS code
with parity check matrix $H_{\mathcal{RS}(n,\delta)}$. The elements in the codeword must be all nonzero.
Then all such codes consist a
subclass of Alternant codes, which we call Sub-Alternant codes. The code in the subclass is denoted by
$\mathcal{S}{\rm-}\mathcal{A}_r(\mathbf{a},\mathbf{y})$.

In this section, we only consider the binary primitive Alternant codes, i.e.,
we take $ q=2,\ n=2^m-1,\ \alpha_i=\alpha^i,\ 0\leq i\leq n-1,\ r=n-k$.
Then the parity check matrix of the binary primitive Alternant code $\mathcal{A}_r(\mathbf{a}, \mathbf{y})$ is given by
\begin{equation}
\label{parity check matrix of binary primitive alternant codes}
H_{\mathcal{A}_r(\mathbf{a}, \mathbf{y})} =
\left(
\begin{array}{cccc}
y_1&y_2&\cdots&y_n\\
y_1&y_2\alpha&\cdots&y_n\alpha^{(n-1)}\\
\vdots&\vdots&\vdots&\vdots\\
y_1&y_2\alpha^{r-1}&\cdots&y_n\alpha^{(n-1)(r-1)}
\end{array}
\right).
\end{equation}
It is easy to see that $H_{\mathcal{A}_r(\mathbf{a}, \mathbf{y})}=H_{\mathcal{RS}(n,r+1)}\cdot {\rm diag}(\mathbf{y})$ where ${\rm diag}(\mathbf{y})$ is a diagonal matrix with $\mathbf{y}$ as
the diagonal elements.
\begin{definition}
\label{definition of SA code}
For any
$\mathbf{y}=(y_1, y_2, \ldots, y_n)\in \mathcal{RS}(n,\delta)$
whose every position is nonzero element, i.e., $H_{\mathcal{RS}(n,\delta)}\mathbf{y}^T=0$, and
$y_i\neq 0$ for all $1\leq i\leq n$. Then $\mathcal{S}{\rm -}\mathcal{A}_r(\mathbf{a},\mathbf{y})$ is defined as:
\[
\mathcal{S}{\rm -}\mathcal{A}_r(\mathbf{a},\mathbf{y})=\{c\in \mathbb{F}_2^n|H_{\mathcal{A}_r(\mathbf{a}, \mathbf{y})}c^T=0 \}
\]
where $H_{\mathcal{RS}(n,\delta)}$ is the parity check matrix in (\ref{parity check matrix of general RS codes}) and $H_{\mathcal{A}_r(\mathbf{a}, \mathbf{y})}$ is the one in (\ref{parity check matrix of binary primitive alternant codes}).
\end{definition}

We have the following asymptotic behavior of  these Sub-Alternant codes.
\begin{lemma}
\label{asymptotically good SubAlternant codes}
Let $\delta/2< r<\min\{\delta,n/2\}$, there exist long codes $\mathcal{S}{\rm -}\mathcal{A}_r(\mathbf{a},\mathbf{y})$ 
meeting the
Gilbert-Varshamov bound.
\end{lemma}
\begin{IEEEproof}
Consider any binary word $\mathbf{c}=(c_1,c_2,\ldots,c_n)$ of weight $t$. For $\mathbf{c}$ to be a codeword of $\mathcal{S}{\rm -}\mathcal{A}_r(\mathbf{a},\mathbf{y})$,
it must satisfy $H_{\mathcal{A}_r(\mathbf{a},\mathbf{y})}\mathbf{c}^T=0$. Then
\[H_{\mathcal{RS}(n,r+1)}(y_1c_1,y_2c_2,\ldots,y_nc_n)^T=0.\]
Let the nonzero elements in $\mathbf{c}$ be
$\{c_{i_1}, c_{i_2}, \ldots, c_{i_t}\}$ where $1\leq i_1<i_2<\cdots<i_t\leq n$. Then we have
\[H_{\mathcal{RS}(n,r+1)}(\ldots,y_{i_1}c_{i_1},\ldots,y_{i_t}c_{i_t},\ldots)^T=0,\] where ``$\ldots$" denote the
zero elements if necessary. This implies that $H_{\mathcal{RS}(n,r+1)}(\ldots,y_{i_1},\ldots,y_{i_t},\ldots)^T=0$
because $\mathbf{c}$ is binary. If we let
\[
B^{'}_w = (2^m-1)\sum_{j=0}^{w-(r+1)}(-1)^j \binom{w-1}{j}2^{m(w-(r+1)-j)},
\]
then the Hamming weight
enumerator of the RS code with parity check matrix $H_{\mathcal{RS}(n,r+1)}$
is $B_w = \binom{n}{w}B^{'}_w$. Then the number of $(\ldots,y_{i_1},\ldots,y_{i_t},\ldots)$ is at most $B^{'}_t$.

According to Definition \ref{definition of SA code} and $r<\delta$, we have
$H_{\mathcal{RS}(n,r+1)}(y_1,y_2,\ldots,y_n)^T=0$. Then
\[
H_{\mathcal{RS}(n,r+1)}(\ldots,y_{j_1},\ldots,y_{j_{(n-t)}},\ldots)^T=0,
\]
where $(\ldots, y_{j_1}, \ldots, y_{j_{(n-t)}}, \ldots)^T=(y_1,y_2,\ldots,y_n)^T-(\ldots,y_{i_1}, \ldots,y_{i_t},\ldots)^T$,
 $1\leq j_1<j_2<\ldots<y_{j_{(n-t)}}\leq n$,
``$\ldots$" denote the zero elements if necessary.
Then the number of $(\ldots, y_{j_1}, \ldots, y_{j_{(n-t)}}, \ldots)$ is at most $B^{'}_{n-t}$.
Therefore the number of  $\mathbf{y}=(y_1,y_2,\ldots,y_n)$ is at most $B^{'}_tB^{'}_{n-t}$. Notice that
\[
B^{'}_w\leq (2^m-1)^{w-r} ,
\]then
\[
B^{'}_tB^{'}_{n-t}\leq(2^m-1)^{n-2r}.
\]
Therefore for all codewords of weight $t<\omega$, the number of vectors $\mathbf{y}$ that
include such codewords in the corresponding Alternant code
$\mathcal{S}$-$\mathcal{A}(\mathbf{a},\mathbf{y})$ is at most
\[
\sum_{t=r+1}^{\omega-1}B^{'}_tB^{'}_{n-t}\binom{n}{t}\leq (2^m-1)^{n-2r}\sum_{t=r+1}^{\omega-1}\binom{n}{t}.
\]

On the other hand, the total number of such Alternant codes equal to the number of choices for
$\mathbf{y}$, which is
\begin{eqnarray*}
A_n &=&(2^m-1)\sum_{j=0}^{n-\delta}(-1)^j \binom{n-1}{j}2^{m(n-\delta-j)}\\
&\geq&(2^m-1)2^{m(n-\delta)}(1-\frac{n-1}{2^m}) \\
&>&(2^m-1)^{n-\delta}.
\end{eqnarray*}
So if
\[
(2^m-1)^{n-2r}\sum_{t=r+1}^{\omega-1}\binom{n}{t}<
(2^m-1)^{n-\delta}
\]
which can be simplified
\[
\sum_{t=r+1}^{\omega-1}\binom{n}{t}<(2^m-1)^{2r-\delta},
\]
there exists a $[2^m,\geq2^m-m(2r-\delta),\geq\omega]$ code.
Using the estimates of binomial coefficients in \cite[Ch.10. Corollary 9]{macwilliams1977theory} and taking the limit as $n\rightarrow \infty$, we can write this condition as
\begin{equation}
H(\frac{d}{n})+o(1)<\frac{m(2r-\delta)}{n}+o(1).
\end{equation}

Let $\tau=2r-\delta$, $\epsilon=o(1)$ and choose the values of parameters properly, then there exists a Sub-Alternant code with $m\tau/n=H(d/n)+\epsilon$.
And by a property of Alternant codes, the rate $R$ of this code satisfies
\begin{eqnarray}
R&\geq&1-\frac{m\tau}{n}\nonumber\\
&>&1-H(\frac{d}{n})-\epsilon.
\end{eqnarray}

Hence the above Sub-Alternant code is asymptotically close to the Gilbert-Varshamov bound.
\end{IEEEproof}

From Definition \ref{definition of SA code} and Lemma \ref{asymptotically good SubAlternant codes},
we have the following result directly.
\begin{theorem}
\label{asymptotically good quantum alternant codes}
There exists a family of AQCs with parameters
\[
[[n,\geq n-mr-1,\geq r+1/ 2]]
\]
where $3\leq n\leq 2^m+1$, $1<r<\delta<n$.

As $n\rightarrow \infty$ and $\delta/2< r<\min\{\delta, n/2\}$, there exist a family $\mathcal{Q}$ of
asymptotically $Z$-type good AQCs  such that
\[
R_\mathcal{Q}=1-H(\delta_z)-\epsilon,
\]
\[
\delta_x=\frac{2}{n}\rightarrow 0,
\]
\[
0<\delta_z<\frac{1}{2}.
\]
\end{theorem}
\begin{IEEEproof}
Let $I=\underbrace{[1\;1\;\cdots\;1]}_{n}$ and $C_1=[n,n-1,2]$ with
$I$ as its parity check matrix. For any
$C_2=\mathcal{S}{\rm -}\mathcal{A}_r(\mathbf{a},\mathbf{y})$ and let $r<\delta$,
 we have
\begin{eqnarray*}
H_{\mathcal{A}(\mathbf{a}, \mathbf{y})}\cdot I^T&=&H_{\mathcal{RS}(n,r+1)}
\cdot {\rm diag}(\mathbf{y})\cdot I^T\\
&=&H_{\mathcal{RS}(n,r+1)}\cdot\mathbf{y}^T\\
&=&0.
\end{eqnarray*}
Therefore $C_1^\bot\subseteq C_2$. By Lemma \ref{AQC Constructions} there exists a
family of
 AQCs with parameters
\[
[[n,\geq n-mr-1,\geq r+1/ 2]]_q
\]
where $3\leq n\leq q^m+1$, $1<r<\delta<n$.

The asymptotic result follows  from Lemma \ref{asymptotically good SubAlternant
codes} immediately.
\end{IEEEproof}

It shows that when $d_x = 2$,  $Z$-parts of
our new AQCs can attain the classical Gilbert-Varshamov bound,
not just the quantum version.
\begin{table*}[!t]
\renewcommand{\arraystretch}{1.2}
\caption{Good Binary AQCs constructed from nested Goppa codes using Magma}
\label{computed nested Goppa codes using Magma}
\centering
\begin{tabular}[c]{|c|clc|lcl|}
\hline
No.&Field&$\Gamma(L,G)$&$G(z)$&$\Gamma(L,F)^\bot$&$F(z)$&$[[n,k,d_z/d_x]]$\\
\hline
$1$&$\mathbb{F}_{2^6}$&$[55,16, 19]$(OPC)&$z^{9}+1$&$[55,49,3]$(OPC)&$(z-1)^{6}\cdot G(z)$&
$[[55,10, 19/3]]$\\

$2$&$\mathbb{F}_{2^6}$&$[56,16, 20]$(OPC)&ETC&$[56,50,3]$(OPC)&DETC&
$[[56,10, 20/3]]$\\

$3$&$\mathbb{F}_{2^6}$&$[54,16,18]$(OPC)&PTC&$[54,48,3]$(OPC)&DPTC&
$[[54,10,18/3]]$ \\

$4$&$\mathbb{F}_{2^6}$& $[55,16, 19]$(OPC)
 &$z^{9}+1$&$[55,45,4]$(BKLC)&$(z-1)^{2}\cdot G(z)$&
$[[55,6, 19/4]]$ \\

$5$&$\mathbb{F}_{2^6}$&$[55,15, 20]$(OPC)&EPC&$[55,46,3(4)]$&DEPC&
$[[55,6, 20/3]]$  \\

$6$&$\mathbb{F}_{2^6}$&$[56,16, 20]$(OPC)&ETC&$[56,46,4]$(BKLC)&DETC&
$[[56,6, 20/4]]$  \\

$7$&$\mathbb{F}_{2^6}$&$[54,15,19]$(OPC)&STC&$[54,45,3(4)]$&DSTC&
$[[54,6, 19/3]]$ \\

$8$&$\mathbb{F}_{2^6}$&$[54,16,18]$(OPC)&PTC&$[54,44,4]$(BKLC)&DPTC&
$[[54,6,18/4]]$ \\

\hline
$9$&$\mathbb{F}_{2^8}$&$[239,123, 35]$(OPC)&$z^{17}+1$&$[239,229,4]$(BKLC)&$(z-1)^{60}\cdot G(z)$&$[[239,113, 35/4]]$\\

$10$&$\mathbb{F}_{2^8}$&$[239,122, 36]$(OPC)&EPC&$[239,230,3(4)]$&DEPC&
$[[239,113, 36/3]]$\\

$11$&$\mathbb{F}_{2^8}$&$[240,123, 36]$(OPC)&ETC&$[240,230,4]$(BKLC)&DETC&$[[240,113, 36/4]]$\\

$12$&$\mathbb{F}_{2^8}$&$[238,122,35]$(OPC)&STC&$[238,229,3(4)]$&DSTC&$[[238,113, 35/3]]$ \\
$13$&$\mathbb{F}_{2^8}$&$[238,123,34]$(OPC)&PTC&$[238,228,4]$(BKLC)&DPTC&
$[[238,113,34/4]]$ \\

$14$&$\mathbb{F}_{2^8}$&$[239,123, 35]$(OPC)&$z^{17}+1$&$[239,218,6]$(BKLC)&$(G(z))^5$&$[[239,102, 35/6]]$   \\

$15$&$\mathbb{F}_{2^8}$&$[239,122, 36]$(OPC)&EPC&$[239,219,5(6)]$&DEPC&
$[[239,102, 36/5]]$ \\

$16$&$\mathbb{F}_{2^8}$&$[240,123, 36]$(OPC)&ETC&$[238,217,6]$(BKLC)&DETC&
$[[238,102,34/6]]$ \\

$17$&$\mathbb{F}_{2^8}$&$[238,122,35]$(OPC)&STC&$[240,219,6]$(BKLC)&DSTC&
$[[240,102, 36/6]]$ \\

$18$&$\mathbb{F}_{2^8}$&$[238,123,34]$(OPC)&PTC&$[238,218,5(6)]$&DPTC&
$[[238,102, 35/5]]$\\

$19$&$\mathbb{F}_{2^8}$&$[239,123, 35]$(OPC)&$z^{17}+1$&$[239,208,8]$(BKLC)&$(z-1)^{30}\cdot G(z)$&$[[239,92, 35/8]]$\\

$20$&$\mathbb{F}_{2^8}$&$[239,122, 36]$(OPC)&EPC&$[239,209,7(8)]$&DEPC&
$[[239,92, 36/7]]$\\

$21$&$\mathbb{F}_{2^8}$&$[240,123, 36]$(OPC)&ETC&$[240,209,8]$(BKLC)&DETC&
$[[240,92, 36/8]]$ \\

$22$&$\mathbb{F}_{2^8}$&$[238,122,35]$(OPC)&STC&$[238,208,7(8)]$&DSTC&
$[[238,92, 35/7]]$\\

$23$&$\mathbb{F}_{2^8}$&$[238,123,34]$(OPC)&PTC&$[238,207,8]$(BKLC)&DPTC&
$[[238,92,34/8]]$ \\

\hline

\end{tabular}
\end{table*}

\section{AQCs From Nested Goppa Codes}

In 1970s, V. D. Goppa introduced a class of linear codes
called Goppa codes or $\Gamma(L,G)$ codes which form an important subclass of Alternant codes and
asymptotically meet the Gilbert-Varshamov bound \cite{macwilliams1977theory}.
\begin{definition}
Let $G(z)$ be a monic polynomial
with coefficients from $\mathbb{F}_{q^m}$, $L=\{\alpha_1,\alpha_2,\ldots,\alpha_n\}\subseteq\mathbb{F}_{q^m}[z]$ such that
$\forall i,G(\alpha_i)\neq 0$. The Goppa code $\Gamma(L,G)$ of length $n$ over $\mathbb{F}_q$, is the set of codewords $c=(c_1,c_2,\ldots,c_n)\in \mathbb{F}_q^n$ such that
\begin{equation}
\label{definition of Goppa codes}
\sum_{i=1}^n\frac{c_i}{z-\alpha_i} = 0\ {\rm mod}\ G(z)
\end{equation}
\end{definition}
$G(z)$ is called the Goppa polynomial, $L$ is the location set.

We have the following nested Goppa codes which are similar to nested cyclic codes.
\begin{lemma}
\label{nested Goppa codes}
Let $G(z)$, $F(z)$ be  Goppa polynomials of $q$-ary Goppa codes $\Gamma(L,G)$ and $\Gamma(L,F)$ respectively. If $F(z) | G(z)$,
then $\Gamma(L,G) \subseteq \Gamma(L,F)$.
\end{lemma}
\begin{IEEEproof}
Let $G(z)\in \mathbb{F}_{q^m}[z]$ be a monic polynomial of degree $r_1$. Then we can decompose the Goppa polynomial $G(z)$ into distinct irreducible polynomials $G_u(z)$ over $\mathbb{F}_{q^m}$ as: $G(z)=\prod_{u=1}^s\{G_u(z)\}^{d_u}$, where $d_u$ and s are integers that satisfy $\sum_{u=1}^sd_u(\deg G_u(z))=r_1$, $\deg G_u(z)\geq1$.
Since the polynomials $G_u(z)$, $u=1,2,\ldots,s$ are relatively prime, the defining set (\ref{definition of Goppa codes}) for $\Gamma(L,G)$ can be rewritten as:
\begin{equation}
\label{redefintion of Goppa codes}
\sum_{i=1}^n\frac{c_i}{z-\alpha_i} = 0\ {\rm mod}\ \{G_u(z)\}^{d_u},
\end{equation}
for $u=1,2,\ldots,s$. (\ref{definition of Goppa codes}) and (\ref{redefintion of Goppa codes}) are equivalent for $\Gamma(L,G)$.

Since $F(z)|G(z)$,  then:
\[
F(z)=\prod_{v\in\{u_1,\ldots,u_t\}}\{G_v(z)\}^{f_v}
\]
where $t$ and $f_v$ are integers, and  $\{u_1,u_2,\ldots,u_t\}\subseteq\{1$, $2,\ldots,s\}$,  $0\leq f_v\leq d_v$, $v\in\{u_1,u_2,\ldots,u_t\}$.

It is easy to see that, for every $c=(c_1,c_2,\ldots,c_n)\in\Gamma(L,G)$ which satisfies (\ref{redefintion of Goppa codes}) also satisfies
\begin{equation*}
\sum_{i=1}^n\frac{c_i}{z-\alpha_i} = 0\ {\rm mod}\ \{G_v(z)\}^{f_v},
\end{equation*}
for  $v=u_1,u_2,\ldots,u_t$.

Then, there is $c=(c_1,c_2,\ldots,c_n)\in\Gamma(L,F)$. Therefore
$\Gamma(L,G) \subseteq \Gamma(L,F)$
\end{IEEEproof}

From Lemma \ref{nested Goppa codes}, we know that the nested Goppa codes are widespread. People have found that certain Goppa codes
have good properties and some of these codes have the best
known minimum distance of any known codes with the same
length and rate. It induces us to identify these codes and investigate their nested relationship. And we use Magma to compute the dual distance  of nested Goppa codes to some computationally reasonable length. Some  good AQCs are given in TABLE \ref{computed nested Goppa codes using Magma}. The shorthands in the tables are explained as follows. If a code is both BKLC and BDLC, or achieves the upper bound, we call it OPC(optimal code).
``EPC" stands for expurgated code, ``ETC" stands for extended code, ``STC" stands for shortened code and ``PTC" stands for punctured code. ``DEPC" stands for the dual of expurgated code, others are the same. ``$d=3(4)$", for example, means the minimum distance is 3, and the corresponding BKLC's distance is 4. ``Dim" stands for dimension of the code. ``LB" stands for lower bound of the code. Firstly we give an explicit example below.
\begin{example}
 \label{Goppa55}
Loeloeian and Conan gave a $\Gamma(L,G)=[55,16,19]$ binary Goppa code in \cite{loeloeian198455} which is  a BKLC (Best known linear code), a BDLC (Best dimension linear code) and a BLLC (Best length linear code) over $\mathbb{F}_2$ in the databases of Magma and \cite{Grassl:codetables}. The Goppa polynomial of $\Gamma(L,G)$ is given by
 \begin{eqnarray*}
 G(z)=(z-\alpha^9)(z-\alpha^{12})(z-\alpha^{30})(z-\alpha^{34})(z-\alpha^{42})\\
 \cdot(z-\alpha^{43})(z-\alpha^{50})(z-\alpha^{54})\hspace*{-7.5mm}
\end{eqnarray*}
where $\alpha$ is a primitive element of $\mathbb{F}_{2^6}$.
Take $\Gamma(L,F)$ with Goppa polynomial $F(z)=(z-\alpha^9)^2\cdot G(z)$, then $\Gamma(L,F)\subseteq\Gamma(L,G)$. Using Magma, we know that $\Gamma(L,F)^\bot=[55,45,4]$.
Then we get an $[[55,6,19/4]]$ AQC.
If $F(z)=(z-\alpha^9)^6\cdot G(z)$, then $\Gamma(L,F)^\bot=[55,49,3]$, we get an $[[55,10,19/3]]$ AQC.
From Theorem \ref{nested Goppa codes of Bezzateev} below, we get an $[[55,15,19/2]]$ AQC. From the databases, we know that $[55,45,4]$, $[55,49,3]$ and $[55,54,2]$ are all BKLCs. $[55,49,3]$ and $[55,54,2]$ are BDLCs and BLLCs as well. Therefore $[[55,10,19/3]]$ and
$[[55,15,19/2]]$ are BDAQCs(Best dimension asymmetric quantum code).
 \end{example}


\begin{table}[!t]
\renewcommand{\arraystretch}{1.2}
\setlength{\tabcolsep}{2pt}
\caption{Binary AQCs constructed from Goppa codes with $d_x=2$}
\label{AQCs with dx=2 from Goppa codes}
\centering
\begin{tabular}[c]{|c|cccccclc|}
\hline
$m$&$t$&$S$&$n$&$G(z)$&Dim&LB&$[[n,k,d_z/d_x]]$&Refs.\\
\hline
 $6$&$3$&$0$&$60$ &$z^3+1$&$43$&$43$&$[[60,42,6/2]]$&\\
 &$7$&$0$&$56$ &$z^7+1$&$17$&$15$&$[[56,16,14/2]]$&\cite{veron2005proof}\\
 &$9$&$1$&$55$ &$z^9+1$&$16$&$1$&$[[55,15,19/2]]$&\cite{SheBezMir89,veron2005proof}\\
 &\normalsize--&\normalsize--&56 &ETC&$16$&\normalsize--&$[[56,15,20/2]]$&\\
 &\normalsize--&\normalsize--&54 &PTC&$16$&\normalsize--&
 $[[54,15,18/2]]$&\\
\hline
 $8$ &$3$&$0$&$252$ &$z^3+1$&$229$&$229$&$[[252,228,6/2]]$&\\
 &$5$&$1$&$251$ &$z^5+1$&$211$&$211$&$[[251,210,11/2]]$&\\
 &\normalsize--&\normalsize--&252 &ETC&$211$&\normalsize--&$[[252,210,12/2]]$&\\
 &\normalsize--&\normalsize--&250 &PTC&$211$&\normalsize--&
 $[[250,210,10/2]]$&\\
 &$15$&$0$&$240$ &$z^{15}+1$&$124$&$121$&$[[240,123,30 /2]]$&\cite{veron2005proof}\\
 &$17$&$1$&$239$ &$z^{17}+1$&$123$&$103$&$[[239,122,35 /2]]$&\cite{SheBezMir89,veron2005proof}\\
 &\normalsize--&\normalsize--&240 &ETC&$123$&\normalsize--&$[[240,122,36/2]]$&\\
 &\normalsize--&\normalsize--&238 &PTC&$123$&\normalsize--&
 $[[238,122,34/2]]$&\\
 &$51$&$0$&$204$ &$z^{51}+1$&$2$&-$203$&$[[204,1,102/2]]$&\\
 \hline
 $9$ &$73$&$1$&$439$ &$z^{73}+1$&$58$&-$218$&$[[439,57,147/2]]$&\cite{bezzateev1998subclass}\\
 &\normalsize--&\normalsize--&440 &ETC&$58$&\normalsize--&$[[440,57,148/2]]$&\\
 &\normalsize--&\normalsize--&438 &PTC&$58$&\normalsize--&$[[438,57,146/2]]$&\\
 \hline
 $10$ &$31$&$0$&$992$ &$z^{31}+1$&$687$&$683$&$[[992,686, 62/2]]$&\cite{veron2005proof}\\
   &$33$&$1$&$991$ &$z^{33}+1$&$686$&$661$&$[[991,685,  67/2]]$&\cite{SheBezMir89,veron2005proof}\\
   &\normalsize--&\normalsize--&$992$ &ETC&$686$&\normalsize--&$[[992,685,68/2]]$&\\
     &\normalsize--&\normalsize--&$990$ &PTC&$686$&\normalsize--&$[[990,685,66/2]]$&\\
   &$93$&$1$&$931$ &$z^{93}+1$&$105$&$1$&$[[931,104,  187/2]]$&\\
   &\normalsize--&\normalsize--&$932$ &ETC&$105$&\normalsize--&$[[932,104,188/2]]$&\\
     &\normalsize--&\normalsize--&$930$ &PTC&$105$&\normalsize--&$[[930,104,186/2]]$&\\
\hline
 $11$&$89$&$1$&$1959$ &$z^{89}+1$&$980$&$980$&$[[1959,979, 179/2]]$&\\
  &\normalsize--&\normalsize--&$1960$ &ETC&$979$&\normalsize--&$[[1960,979,180/2]]$&\\
 &\normalsize--&\normalsize--&$1958$ &PTC&$979$&\normalsize--&$[[1958,979,178/2]]$&\\
\hline
$12$&$63$&$0$&$4032$ &$z^{63}+1$&$3282$&$3277$&$[[4032,3281,126/2]]$&\cite{veron2005proof}\\

&$65$&$1$&$4031$ &$z^{65}+1$&$3281$&$3251$&$[[4031,3280,131/2]]$&\cite{SheBezMir89,veron2005proof}\\
  &\normalsize--&\normalsize--&$4032$ &ETC&$3281$&\normalsize--&$[[4032,3280,132/2]]$&\\
 &\normalsize--&\normalsize--&$4030$ &PTC&$3281$&\normalsize--&$[[4030,3280,130/2]]$&\\

  &$195$&$0$&$3900$ &$z^{195}+1$&$1759$&$1561$&$[[3900,1758,390/2]]$&\\

&$273$&$1$&$3823$ &$z^{273}+1$&$1311$&$547$&$[[3823,1310, 547/2]]$&\cite{bezzateev1998subclass}\\
 &\normalsize--&\normalsize--&$3824$ &ETC&$1311$&\normalsize--&$[[3824,1310,548/2]]$&\\
 &\normalsize--&\normalsize--&$3822$ &PTC&$1311$&\normalsize--&$[[3822,1310,546/2]]$&\\
   &$315$&$0$&$3780$ &$z^{315}+1$&$474$&$1$&$[[3780,473,630/2]]$& \\
   &$455$&$0$&$3640$ &$z^{455}+1$&$197$&-$1819$&$[[3640,196,910/2]]$&\\

  &$585$&$1$&$3511$ &$z^{585}+1$&$196$&-$3509$&$[[3511,195, 1171/2]]$&\\
 &\normalsize--&\normalsize--&$3512$ &ETC&$196$&\normalsize--&$[[3512,195,1172/2]]$&\\
 &\normalsize--&\normalsize--&$3510$ &PTC&$196$&\normalsize--&$[[3510,195,1170/2]]$&\\
  &$819$&$0$&$3276$ &$z^{819}+1$&$2$&-$6551$&$[[3276,1,1638/2]]$&\\

\hline
\end{tabular}
\end{table}

In \cite{bezzateev1995subclass}, Bezzateev and Shekhunova described a subclass of
Goppa codes with minimal distance equal to the design distance. We find that  their codes can  be used to construct AQCs with $d_x=2$.
\begin{theorem}
\label{nested Goppa codes of Bezzateev}
Let the polynomial $\mathcal{G}(z)=z^t+A\in \mathbb{F}_{2^m}[z]$, where $t|(2^m-1)$, i.e., $2^m-1=t\cdot l$ and $A$ is a $t$th power in $\mathbb{F}_{2^m}\backslash \{0\}$.  $\mathcal{N} =\{ \alpha \in \mathbb{F}_{2^m}:\mathcal{G}(\alpha)\neq 0\}$. Denote $S=\sum_{\mu=1}^{l-1}1/(\alpha^{\mu t}+1)$, $\alpha$ is a primitive element of $\mathbb{F}_{2^m}$. Then $S$ must be $1$ or $0$.
\begin{itemize}[\IEEEsetlabelwidth{2)}]
\item[(1)]  If $S=1$, then for a Goppa code $\Gamma(L,G)$ with Goppa polynomial $G(z)=\mathcal{G}(z)$
and $ L=\mathcal{N}$,
there exists an AQC  with parameters
\[
[[2^m-t,\geq 2^m-t-mt-1,2t+1/2]],
\]
this code can be extended to
\[
[[2^m-t+1,\geq 2^m-t-mt-1,2t+2/2]],
\]
and can be punctured to
\[
[[2^m-t-1,\geq 2^m-t-mt-1,2t/2]].
\]
\item[(2)]  If $S=0$, for punctured $\Gamma(L,G)$ with $G(z)=\mathcal{G}(z)$
and $ L=\mathcal{N}-\{0\}$,
there exists a  punctured AQC  with parameters
\[
[[2^m-t-1,\geq 2^m-t-mt-1,\geq 2t/2]].
\]
\end{itemize}
\end{theorem}
\begin{IEEEproof}
We follow the proof process of Theorem 2.1 given by Bezzateev \& Shekhunova in
\cite{bezzateev1995subclass}. For simplicity, we take $A=1$.
For $S=\sum_{\mu=1}^{l-1}1/(\alpha^{\mu t}+1)$, then $S=1$ or $0$ as $S=S^2$.

(1) If $S=1$. We take $G(z)=\mathcal{G}(z)=z^t+1$,
$L=\mathcal{N}=\{\alpha_1,\alpha_2,\ldots,\alpha_n\}$. For $1\leq \mu\leq l-1$, we
consider   binary vectors $\mathbf{a}_\mu=(a_1^\mu,a_2^\mu,\ldots,a_n^\mu)$ with
Hamming weight $t$ and such that its nonzero components are on positions which
correspond to the following subset of $L$:
\[
\{(\alpha^l)^i\cdot\beta_\mu,\ \ i=0,1,\ldots,t-1\}
\]
$\alpha$ is a primitive element of $\mathbb{F}_{2^m}$ and $\beta_\mu=\alpha^\mu$.
Then
\[
\sum_{j=1}^{n}a_j^\mu\frac{1}{x-\alpha_j}=\frac{1}{\beta_\mu^t+1}x^{t-1} \mod x^t+1
\]
for $1\leq \mu\leq l-1$.

Let the last binary vector $\mathbf{a}_l=(a_1^l,a_2^l,\ldots,a_n^l)$ have only one
nonzero component on the position which correspond to $\{0\}$. Then for this vector
\[
\sum_{j=1}^{n}a_j^l\frac{1}{x-\alpha_j}=x^{t-1} \mod x^t+1.
\]

Now let us consider the sum of vectors $\mathbf{a}_1, \mathbf{a}_2,
\ldots,\mathbf{a}_l$
\begin{eqnarray*}
\sum_{j=1}^{n}\sum_{\mu=1}^la_j^\mu\frac{1}{x-\alpha_j}=
(\frac{1}{\beta_1^t+1}+
\cdots+\frac{1}{\beta_{l-1}^t+1}+1)\\
\cdot x^{t-1} \mod x^t+1.\hspace*{-8mm}
\end{eqnarray*}
So as
$S=\sum_{\mu=1}^{l-1}\frac{1}{\beta_\mu^t+1}=\sum_{\mu=1}^{l-1}\frac{1}{\alpha^{\mu
t}+1}=1$, then
\[
\sum_{j=1}^{n}\sum_{\mu=1}^la_j^\mu\frac{1}{x-\alpha_j}=
0 \mod x^t+1.
\]
Thus vector
$\mathbf{a}=\mathbf{a}_1+\mathbf{a}_2+\cdots+\mathbf{a}_l=(1,1,\ldots,1)$ is a
codeword  of the Goppa polynomial $G(z)=z^t+1$ and $L=\mathcal{N}$ and its Hamming
weight is equal to $2^m-t$. Therefore 
there exists an AQC  with parameters
\[
[[2^m-t,\geq 2^m-t-mt-1,2t+1/2]],
\]
this code can be extended into
\[
[[2^m-t+1,\geq 2^m-t-mt-1,2t+2/2]],
\]
and can be punctured into
\[
[[2^m-t-1,\geq 2^m-t-mt-1,2t/2]].
\]

(2) If $S=0$, we take $\Gamma(L,G)$ with $G(z)=\mathcal{G}(z)$
and $ L=\mathcal{N}-\{0\}$, the proof is similar to (1) above.
And we can omit  the last binary vector $\mathbf{a}_l=(a_1^l,a_2^l,\ldots,a_n^l)$
as $S=0$. Then
there exists a  punctured AQC  with parameters
\[
[[2^m-t-1,\geq 2^m-t-mt-1,\geq 2t/2]].
\]
\end{IEEEproof}

From the proof of Theorem
\ref{nested Goppa codes of Bezzateev}, we know that classical codes corresponding to $X$-parts of AQCs are all $[n, n-1, 2]$ optimal codes. Therefore the error correction abilities of the corresponding Goppa codes are all transformed into $Z$-parts of AQCs with only one information bit loss each.
 Maatouk $et\ al.$ \cite{maatouk2007good} found  that the classical codes described in Theorem \ref{nested Goppa codes of Bezzateev} achieved better than the GV bound when the field size is small.
For some ``typical" cases, the estimation of the dimension is much better than the lower bound \cite{veron1998goppa,SheBezMir89,bezzateev1998subclass}, and sometimes   the estimation is the true dimension \cite{veron2001true,veron2005proof}. AQCs derived from Theorem \ref{nested Goppa codes of Bezzateev} are given in TABLE \ref{AQCs with dx=2 from Goppa codes}. When the field size is large we only give partial AQCs with loose lower bound(LB).

\section{Asymptotically Good  Binary Expansion of Quantum GRS Codes }
In \cite{retter1991average}, Retter showed that most
binary expansions of GRS codes are asymptotically good.
\begin{theorem}[{\cite[Theorem 1]{retter1991average}}]
\label{theorem of retter1991}
For any small $\epsilon>0$, there exists an $n$ such that the binary
expansions of most GRS codes of any length greater than $n$ satisfy
\begin{equation*}
H(\frac{d}{n})>1-\frac{k}{n}-\epsilon
\end{equation*}
\end{theorem}

From \cite{ashikhmin2001asymptotically}, we have the following result.
\begin{corollary}
Let $C_1$ and $C_2$ be codes over $\mathbb{F}_{2^m}$ and $C_2^\bot\subseteq C_1$.
Let $\alpha_i, i=1,...,m$, be self-dual basis of $\mathbb{F}_{2^m}$ over $\mathbb{F}_2$, i.e.,
\[
{\rm Tr}(\alpha_i\alpha_j)=\delta_{ij}.
\]
Let $D_1$ and $D_2^\bot$ be codes obtained by the symbolwise binary expansion of
codes $C_1$ and $C_2^\bot$ in the basis $\alpha_i$. Then $D_2^\bot\subseteq D_1$
and $D_2^\bot$ is the binary dual of $C_2$.
\end{corollary}

Let $N=2^m-1, N/2\leq K_1\leq K_2\leq N-1$ be integers, for a GRS code $\mathcal{GRS}_{K_1}(\mathbf{a},\mathbf{v})$ of length $N$.
It follows immediately that
$\mathcal{GRS}_{K_1}(\mathbf{a},\mathbf{v})^\bot=
\mathcal{GRS}_{N-K_1}(\mathbf{a},\mathbf{y})\subseteq \mathcal{GRS}_{K_1}(\mathbf{a},\mathbf{y})
\subseteq \mathcal{GRS}_{K_2}(\mathbf{a},\mathbf{y})$, where
 $y_i\cdot v_i=1/\prod_{j\neq i}(\alpha_j-\alpha_i)=\alpha_i,1\leq i\leq N$.
 Then there exists a corresponding AQC with parameters:
 \begin{equation}
 \label{asymmetric quantum GRS codes}
 [[N,K_1+K_2-N,N-K_1+1/N-K_2+1]]_{2^m}.
 \end{equation}

 Denote $C_1=\mathcal{GRS}_{K_1}(\mathbf{a},\mathbf{v})$ and
 $C_2=\mathcal{GRS}_{K_2}(\mathbf{a},\mathbf{y})$ of length $N$. Then
 $C_2^\bot\subseteq C_1$.
 The binary expansions of $C_1$ and $C_2$
 with respect to a self-dual basis give  $D_2^\bot\subseteq D_1$ of
 binary codes with parameters $n=mN$, $k_1=mK_1$, $k_2=mK_2$.

From Theorem \ref{theorem of retter1991}, we can choose suitable $\mathbf{y}$ to make sure $D_2$ is asymptotically good. Because $y_i\cdot v_i=1/\prod_{j\neq i}(\alpha_j-\alpha_i)=\alpha_i,1\leq i\leq N$, then different $\mathbf{y}$ gives different $\mathbf{v}$. Since the binary
expansions of most GRS codes are asymptotically good when $n$ is large, there always exist  the corresponding $\mathbf{v}$ which also give asymptotically good $D_1$.

Summing up, we have the following theorem.
\begin{theorem}
\label{asymptotically good AQCs from Retter}
For a pair of  $(\alpha_1, \alpha_2)$
real numbers satisfying $0<\alpha_1\leq\alpha_2<1/2$, there
exists a family $\mathcal{Q}$ of AQCs which can attain the asymmetric quantum Gilbert-Varshamov bound with
\begin{equation*}
R_{\mathcal{Q}}=1-\alpha_1-\alpha_2,
\end{equation*}
\begin{equation*}
\delta_x\geq H^{-1}(\alpha_1),
\end{equation*}
\begin{equation*}
\delta_z\geq H^{-1}(\alpha_2).
\end{equation*}
\end{theorem}
\begin{IEEEproof}
For the asymmetric quantum GRS codes (\ref{asymmetric quantum GRS codes}),
it follows from the CSS constructions Lemma \ref{AQC Constructions} and
Theorem \ref{theorem of retter1991} that there exist a family $\mathcal{Q}$ of AQCs
with parameters
\[
[[n, k_1+k_2-n,d_z/d_x]]_2
\]
where
$n=mN,k_1=mK_1,k_2=mK_2$, $d_x\geq d_1$, and $d_z\geq d_2$,
the corresponding classical codes are $D_1=[n,k_1,d_1]_2$ and $D_2=[n,k_2,d_2]_2$
which satisfy
\[
\frac{k_1}{n}=1-\alpha_1,\ \frac{k_2}{n}=1-\alpha_2,
\]
\[
\delta_1=\frac{d_1}{n}\geq H^{-1}(\alpha_1),
\]
\[
\delta_2=\frac{d_2}{n}\geq H^{-1}(\alpha_2).
\]

Then we have
\[
R_\mathcal{Q}=\frac{k_1}{n}+\frac{k_2}{n}-1=1-\alpha_1-\alpha_2,
\]
\[
\delta_x=\frac{d_x}{n}\geq\delta_1\geq H^{-1}(\alpha_1),
\]
\[
\delta_z=\frac{d_z}{n}\geq \delta_2\geq H^{-1}(\alpha_2).
\]
\end{IEEEproof}

\begin{figure}[!t]
\centering
\includegraphics[width=3.0in]{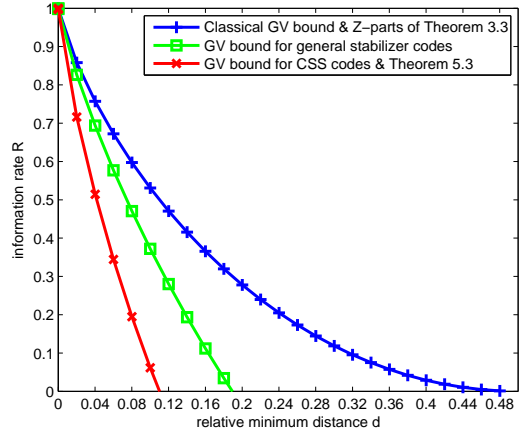}
\caption{Comparison of different versions of binary GV bound.}
\label{GV-Figure}
\end{figure}

Theorem \ref{asymptotically good AQCs from Retter} is also available for QECs. The comparison of classical GV bound and two versions of quantum GV bound is given in Fig. \ref{GV-Figure}.

\section{Conclusion and Discussion}
In this paper, we have constructed several classes of pure asymmetric quantum Alternant codes (AQACs) based on their nested relationships. As a special case, $Z$-parts of our AQACs can attain the classical Gilbert-Varshamov bound when $d_x=2$. We have identified the nested Goppa codes and computed the dual distance of some special Goppa codes. When $d_x=2$, a famous subclass of Goppa codes with fixed minimum distance are converted to AQCs with only one information bit loss each. Some AQACs with good parameters are listed. At last,
Retter's classical results about the asymptotically good binary expansions of   GRS
codes  have been generalized to the quantum situation.

The  asymptotic problem for general AQACs and symmetric quantum Alternant codes is still unsolved. How to construct quantum codes using binary Alternant codes especially binary Goppa codes is an interesting
problem which need further exploring.


%

\ifCLASSOPTIONcaptionsoff
  \newpage
\fi



%
\IEEEoverridecommandlockouts
\bibliographystyle{IEEEtranS}
\bibliography{IEEEabrv,mybibfileQACsConf}

%








\end{document}